\documentclass[%
 reprint,
 showpacs,
 showkeys,
 amsmath,amssymb,
 aps,
 url
]{revtex4-1}

\usepackage{amsmath}
\usepackage{graphicx}
\usepackage{epstopdf}
\usepackage{dcolumn}
\usepackage{bm}
\usepackage{multirow}
\newcommand{\Ref}[1]{(\ref{#1})}
\newcommand{\beao}{\begin{eqnarray*}}
\newcommand{\eeao}{\end{eqnarray*}}
\newcommand{\be}{\begin{equation}}
\newcommand{\ee}{\end{equation}}
\newcommand{\bea}{\begin{eqnarray}}
\newcommand{\eea}{\end{eqnarray}}
\newcommand{\beq}{\begin{eqnarray}}
\newcommand{\eeq}{\end{eqnarray}}
\newcommand{\nn}{\nonumber}

\begin{document}

\title{Estimates of $Z'$~couplings within data on the $A_{FB}$ \\
for Drell-Yan process at the LHC at $\sqrt{s} = 7$ and $8$ TeV }

\author{A.~Pevzner}
 \email{apevzner@omp.dp.ua}
\author{V.~Skalozub}%
 \email{skalozubv@daad-alumni.de}
\affiliation{Oles Honchar Dnipropetrovsk National University, Theoretical Physics Department}%


\date{\today}

\begin{abstract}
Model-independent search for the Abelian $Z'$ gauge boson in the
Drell-Yan  process at the LHC at $\sqrt{s}=7$~ and 8~TeV is fulfilled.
Estimations of the $Z'$ axial-vector coupling $a_f^2$ to the
Standard model fermions,  the couplings of the axial-vector to
lepton vector currents $a_f v_l$  and the couplings of the axial-vector to quark vector currents $a_f v_q$
are derived within data on the
forward-backward asymmetry presented by the CMS Collaboration. The
analysis takes into consideration the behaviour of the differential
cross-section which exhibits itself if the derived already special
relations between the couplings
 proper to the renormalizable theories are accounted for. In
particular, they hold in all the  models of Abelian  $Z'$  usually
considered in the model-dependent analysis of the LHC data.  The
coupling values are estimated at $\sim$92 \% CL by means of the
maximum likelihood function. They  weakly depend on the $Z'$ mass
in the investigated  interval 1.2~TeV~$<m_{Z'}<$~5~TeV.
 Taking into account the dependence of $Z-Z'$ mixing angle $\theta_0$ on $m_{Z'}$ 
 and the LEP constraints $|\theta_0|\sim 10^{-3}-10^{-4}$,
 the optimistic limits on $m_{Z'}$ are established
as $3 < m_{Z'} < 7-8$~TeV. Comparison with the results
of other authors is given.
\end{abstract}



\maketitle

\section{Introduction}
After  discovery of the Higgs boson  at the LHC, the Standard
model (SM) is considered to be  completed. From "practical"
computational point of view it means that the neutral scalar
particle of the mass 125 GeV  has to be taken into consideration
for all the processes investigated. If we also believe that the
spontaneous symmetry breaking mechanism is operating to supply
particle masses, the Higgs particle has to be considered as a
fundamental point like state belonging to a renormalizable theory.
This also concerns new models extending  the SM at high energies
and  containing various scalar particles.

 Searching for new physics is  the main goal of  experiments at the LHC.  One   of  expected heavy particles is an Abelian $Z'$
  gauge boson predicted by numerous extended models (see review papers
   \cite{Leike} -- \cite{ZprEarlyLHC}). It is introduced as the field related with an additional $\tilde{U}(1)$
   group to the SM gauge group. Lower bounds for its mass have been obtained at the LEP
   (\cite{LEP2_1}, \cite{LEP2_2}, \cite{LEP2_3}), Tevatron \cite{Tevatron} and first run LHC
   experiments \cite{ZprEarlyLHC} in either model-dependent or model-independent approaches.
   The present day model-dependent published lower bound on the mass  is $m_{Z'} > 2.5$ TeV from the
    CMS results and $m_{Z'} > 2.9$ TeV from the ATLAS ones. At present about hundred $Z'$ models  are discussed in the literature.
 In model-dependent searches established, only  the most popular ones
 such as $LR$, $ALR$, $\chi$, $\psi$, $\eta$, B - L, $ SSM,$ have been investigated
  and the particle  mass  is estimated. These models are also used as benchmarks in
   introducing the  efficient observables for future experiments at the ILC \cite{Pankov2010}, \cite{Langacker2013}.
 In  this approach, the
    couplings to the SM particles were fixed as in the specific considered models and therefore
     not estimated.  As it also occurred,   the identification reach for different  models    is about the estimated $m_{Z'}$
     lower masses. So that  it is problematic to distinguish the  basic  $Z'$ model at the LHC.
      In such a situation, model-independent approaches are also  very perspective. They give a possibility
      for  estimating not only the particle mass but also some $Z'$ couplings to the SM fermions.
      Hence, definite  classes of the extended models could be restricted.

In  studies  of perspective variables for identification of the
$Z'$ models \cite{Langacker2013},
 in particular, it was  concluded  that, as complementary way, a model-independent approach is very desirable.
    Estimations of couplings can be further used in specifying
     the basic $Z'$ model. Usually, the couplings are considered as independent arbitrary numbers. However, this is not
     the case and they are  correlated parameters, if  some natural requirements,  which this model has to satisfy, are assumed.
In most cases we believe that the basic model is renormalizable
one. Hence, correlations follow and the amount of  free
     parameters  reduces.  Moreover, the correlations between couplings influence kinematics of the processes that gives a possibility for
 introducing the specific observables which uniquely pick out the virtual state of interest --  $Z'$ boson in our case.
  The noted additional requirement  assumes searching for new particles within the class of renormalizable models.
   In other aspects the models are not specified. Below, we  say  "model-independent approach" for the
   analysis
   when either the mass or the couplings must be fitted.  Such type approach is in-between
   the usual
     model-dependent method, when all the couplings are fixed and only the mass $m_{Z'}$ is free parameter, and
      model-independent searches assuming complete independence of couplings describing new
      physics. Recent review
      on searching for the Abelian  $Z'$ boson in the model-independent approach is
      \cite{Gulov2010}

In what follows,  we  search for  the Abelian $Z'$ boson
       belonging to a renormalizable model.  We also assume that there is only one additional heavy
particle relevant at considered energies. There are numerous
models of such type. In particular, most of  $E_6$ motivated
models and mentioned above ones enter  this class.   The used in
the present analysis  relations  \Ref{grgav}  are proper to this
class. In particular, they hold in the models noted above. These
relations have been derived already in two ways \cite{Gulov00a},
\cite{Gulov01}. For convenience of readers, we adduce more details
about them in Appendix B.  In what follows, we say $Z'$ boson for
the  Abelian  one, only.   We also assume
 that the SM is the subgroup of the extended group and therefore no interactions
of the  type  $~Z Z' W^+ W^-$ appear in the  tree-level
Lagrangian.

In the present paper, we search for the $Z'$  at the LHC on the
base of the CMS data on the forward-backward asymmetry, $A_{FB}$,
for the Drell-Yan annihilation process measured  at energy
$\sqrt{s}=7$~TeV \cite{CMS_AFB_7TeV} and 8 TeV \cite{CMS_AFB_8TeV}. As we show below, this
observable is fine sensitive to the $Z'$ signals due to kinematics
properties of the  differential cross-sections of the process.
 The advantage of the Drell-Yan  process is that it is a "pure" one and
 we do not need to take the hadronization effects into consideration. We suppose that
 in this process the $Z'$ manifests itself as the intermediate state like  the $Z$ boson and
  the photon. But it is a heavy particle and all the  loops of it  are decoupled at investigated energies.
  As a result, the $Z'$ exhibits itself as the special kind external field.   It modifies the observables as compared
  to the SM predictions. In paper \cite{CMS_AFB_7TeV} presented by the CMS collaboration  it is  noted that all the
   measured $A_{FB}$ values are in agreement with the SM expectations at $1-2\sigma$ confidence level (CL). So that
 there is no indication of new physics. However, in that data there is a significant number of points closely
 located to the CL area  boundary. So that it is of interest to verify whether the data on $A_{FB}$ could result
  in signals (hints, in fact) for new heavy particle -- the Abelian  $Z'$ gauge boson.

 The $A_{FB}$ of the Drell-Yan lepton-antilepton pair is chosen
as the observable for  the experimental data processing. Reasons
for this  are  discussed  in the next section.   This quantity
turns out to be very sensitive to small  changes of used parameters.
Also, its theoretical uncertainty, which originates from the PDF
uncertainty, is much smaller than the one of the total
cross-sections. Thus, the  $A_{FB}$ yields quite precise results
for  measured quantities.  Also, in recent paper
\cite{FiaschiAccomando} it was motivated the complementarity of
the $A_{FB}$ to the total cross-section in searching for the $Z'$
as resonance state. Our model-independent analysis  supports this
idea  for  lower beam energies. In fact, within huge amount of
data accumulated at the LHC at different energies one is able to
estimate various important parameters which could be used in
further  studies.

As we show,  the CMS data on the $A_{FB}$ admit the $Z'$
existence. By using   the maximum likelihood function  method we
estimate the $Z'$ couplings  to the SM fermions for the $Z'$ mass
in the interval 1.2 TeV $< m_{Z'} < 5 $ TeV and obtain  that these
couplings are to be  non-zero with the 92 \% CL accuracy. Taking into account
the estimated value of $a^2_f$ and experimental upper bound on mixing 
angle  $|\theta_0| \sim 10^{-3}-10^{-4}$ \cite{LEPThetaConstraints09} the estimates of the mass $3 < m_{Z'} < 7-8$~TeV
are derived.

The paper is organized as follows. In the next section we present
the cross-sections of the process investigated and its angular
distributions at various values of the effective mass for lepton
pairs.  The observed behaviour  of different factor functions
entering the cross-section gives reasons for introducing the
$A_{FB}$ as convenient observable. In section 3 the estimations of
the couplings are carried out. Section 4 is devoted to discussion
 and comparison with the results of other
authors. In Appendix A, we present  the  behaviour of the $F_{k}$ factors entering 
Eq. \Ref{eqn:AFBWithZp}. Appendix B contains necessary information about the equations
\Ref{MixingAngle}, \Ref{grgav}. Appendix C includes detailed information about the PDF uncertanties. 

\section{Cross-section with the $Z'$ }

In this section, we calculate the cross-section of the Drell-Yan
process in the model-independent approach and   obtain its
dependence on the $Z'$ couplings.

  We start with the differential cross-section   in the parton model  written in the Collins-Soper frame \cite{C-Sframe}:
\begin{widetext}
\be \label{eqn:PartonLevelXSecInNewVariables}
    \frac{d^3 \sigma}{dM\,dY\,dz} = \sum_{q}{M\left[\mathrm{xf}_q \left(\frac{M}{\sqrt{s}}e^{Y}\right)\mathrm{xf}_{\bar{q}}\left(\frac{M}{\sqrt{s}}e^{-Y}\right)\frac{d \hat{\sigma}_q(z)}{dz}\right.}
    +\left.\mathrm{xf}_q \left(\frac{M}{\sqrt{s}}e^{-Y}\right)\mathrm{xf}_{\bar{q}}\left(\frac{M}{\sqrt{s}}e^{Y}\right)\frac{d \hat{\sigma}_q(-z)}{dz}\right].
\ee
\end{widetext}
Here,  $\hat{\sigma}$ is  the parton-level cross-section:
$\hat{\sigma}_q \equiv \sigma_{q\bar{q} \to l^+ l^-}$ and $ l^+
l^-$ are  final lepton states. Everywhere below we denote the
parton-level quantities with the hatted letters and the
appropriated hadron-level quantities, which are already integrated
with PDFs, with the non-hatted ones. $M$ is dilepton invariant
mass,  $Y$ is an intermediate state rapidity, $z =
\cos\theta_{CS}$, where $\theta_{CS}$ is a dilepton scattering
angle.   We take into account the known relations between the
quark $x_1$ and antiquark $x_2$ momentum fractions: $x_{1,2} =
(M/\sqrt{s})e^{\pm Y}$.   The functions $f_q(x)$ are the PDF distributions,
and the functions $\mathrm{xf}_q(x) = x f_q(x)$ are
pre-implemented in the majority of PDF computer packages.  In
\Ref{eqn:PartonLevelXSecInNewVariables} we sum over the quarks
only, not over  both the quarks and antiquarks.

To proceed we have to calculate  the parton-level cross-section
$\hat{\sigma}_{q\bar{q} \to l^+ l^-}$   taking into account the
$Z'$ contributions. The effective low energy  Lagrangian
describing the interaction of the heavy $Z'$ with the SM particles
was introduced in  \cite{Sirlin}, \cite{Cvetic}. Its part related
to our problem and  describing  interactions between the fermions
and the $Z$ and $Z'$ mass eigenstates reads (see, for example,
\cite{Gulov2010}):
\be\label{ZZplagr}
\begin{split}
{\mathcal{L}}_{Z\bar{f}f}=\frac{1}{2}
Z_\mu\bar{f}\gamma^\mu[(v^\mathrm{SM}_{fZ}+\gamma^5
a^\mathrm{SM}_{fZ})\cos\theta_0 \\
 +(v_f+\gamma^5 a_f)\sin\theta_0]f,
\end{split}
\ee
\be \label{ZZ'plagr}
\begin{split}
{\mathcal{L}}_{Z'\bar{f}f}=\frac{1}{2}
Z'_\mu\bar{f}\gamma^\mu[(v_f+\gamma^5 a_f)\cos\theta_0 \\
-
(v^\mathrm{SM}_{fZ}+\gamma^5a^\mathrm{SM}_{fZ})\sin\theta_0]f,
\end{split}
\ee
where $f$ is an arbitrary SM fermion state; $v^\mathrm{SM}_{fZ}$,
$a^\mathrm{SM}_{fZ}$ are the SM axial-vector and vector couplings
of the $Z$-boson, $a_f$ and $v_f$ are the ones for the $Z'$,
$\theta_0$ is the $Z$--$Z'$ mixing angle. Within the considered
formulation, this angle   is determined by the coupling
$\tilde{Y}_\phi$ of fermions to the scalar field as follows (see
\cite{Gulov2010} and Appendix B for details)
\begin{equation}\label{MixingAngle}
\theta_0 =
\frac{\tilde{g}\sin\theta_W\cos\theta_W}{\sqrt{4\pi\alpha_\mathrm{em}}}
\frac{m^2_Z}{m^2_{Z'}} \tilde{Y}_\phi
+O\left(\frac{m^4_Z}{m^4_{Z'}}\right),
\end{equation}
where $\theta_W$ is the SM Weinberg angle, $\tilde{g}$ is $\tilde{U}(1)$ gauge coupling constant and
$\alpha_\mathrm{em}$ is electromagnetic fine structure
constant. Although the mixing angle is small quantity of order
($m^{2}_{Z}/m^{2}_{Z'}$), it contributes to the $Z$-boson exchange amplitude
and cannot be neglected.

As it  was shown in \cite{Gulov2010}, \cite{Gulov00a},
\cite{Gulov01}, if the extended model is renormalizable and
contains the SM as a subgroup, the relations between the couplings
hold:
\be \label{grgav}
v_f - a_f = v_{f^*} - a_{f^*}, \quad a_f = T_{3f}\tilde{g}\tilde{Y}_\phi.
\ee
Here $f$ and $f^*$ are the partners of the $SU(2)_L$ fermion
doublet ($l^* = \nu_l, \nu^* = l, q^*_u = q_d$ and $q^*_d = q_u$),
$T_{3f}$ is the third component of the weak isospin.
 These relations  are proper for the models of Abelian $Z'$.  They are  just  as in the SM for proper values of the
hypercharges $Y^R_f, Y^L_f, Y_\phi$  of the left-handed, right-handed fermions and scalars. 
 The correlations  can be derived from the necessary requirement of renormalizability  that  there are no new divergent structures  appearing
  in one-loop order.  The divergencies could  appear  at the structures presented  in the initial tree-level Lagrangian, only.
If these conditions do not hold, the theory is not renormalizable.
But if they  fulfil  in one-loop order, there is no  guarantee
that this will be the case in higher orders or with accounting for
of  anomalies. The latter two questions  are more delicate. They
require  detailed information about the particle content of the
model.  Thus, the  correlations \Ref{grgav}  are the necessary conditions for renormalizability.  Other way of derivation \Ref{grgav}  is presented in
Appendix B.

The couplings of the  $Z'$ to the axial-vector fermion current
have a universal absolute value proportional to the $Z'$ coupling
to the scalar doublet. Then, the $Z$--$Z'$ mixing angle
(\ref{MixingAngle}) can be determined by the axial-vector
coupling. As a result, the number of independent parameters is
significantly reduced.  This universality follows due to exchange
of  the scalar particles.  In particular, the relations
\Ref{grgav} hold in Two-Higgs-Doublet SM (see Appendix B). Because
 of the universality, we will omit the subscript $f$  and
write $a$ for axial-vector coupling. It is convenient for what
follows to introduce the "normalized" couplings
\begin{equation} \label{eqn:DimensionlessCouplings}
    \bar{a} = \frac{1}{\sqrt{4\pi}}\frac{m_Z}{m_{Z'}}~a,~~ \bar{v}_f = \frac{1}{\sqrt{4\pi}}\frac{m_Z}{m_{Z'}}~v_f.
\end{equation}

As it follows from (\ref{ZZplagr}), (\ref{ZZ'plagr}), the
Drell-Yan process cross-section has the  contribution from the SM,
the contribution from $Z - Z'$ interference, and  from the $Z'$
part. The last contribution  can be neglected at energies not
close to a $Z'$ resonance peak. Hence, taking into account
(\ref{grgav}), the parton-level cross-section can be written as
\begin{equation} \label{eqn:PartonXSecWithFactors}
\begin{split}
    \frac{d\hat{\sigma_q}}{dz} =\left(\frac{d\hat{\sigma_q}}{dz}\right)_{SM}+ \bar{a}^2 \hat{F}_{q1} + \bar{a}\bar{v}_l \hat{F}_{q2} \\
     + \bar{a}\bar{v}_u \hat{F}_{q3} + \bar{v}_l \bar{v}_u \hat{F}_{q4},
\end{split}
\end{equation}
where $\hat{F}_{qk} = \hat{F}_{qk}(M,z)$ are known from
calculation kinematics  factors,  $q$ in the subscript is "$u$" or
"$d$" (for up and down quarks, respectively), subscript "l"
denotes the $Z'$ to lepton coupling, and subscript "u" denotes the
$Z'$ to up-quark coupling. Thus, there are four unknown parameters
which should be estimated from experiments.  However, due to obvious relation $ a^2 \bar{v}_l \bar{v}_u = \bar{a}\bar{v}_l  \bar{a}\bar{v}_u$  the parameter $ \bar{v}_l \bar{v}_u$  can be expressed through three others. So, in general three-parameter fit is needed.  Let us check whether it is possible to find 
  an integral   observable containing less amount of unknown parameters.

For doing that we  consider the behaviour  of the
$\hat{F}_{qk}$ functions. In our analysis, these functions were
calculated in an improved Born approximation in one-loop order. In
the
 $Z - Z'$ interference  part, the loops with the SM particles
coming from the $Z'$ exchange part were computed
 analytically whereas the SM
contributions have been calculated by using the PYTHIA package.

We investigate the behaviour
of the hadron-level factors
\begin{widetext}
\begin{equation} \label{eqn:HadronLevelZpFactors}
    F_k(M,Y,z) = \sum_{q}{M\left[\mathrm{xf}_q \left(\frac{M}{\sqrt{s}}e^{Y}\right)\mathrm{xf}_{\bar{q}}\left(\frac{M}{\sqrt{s}}e^{-Y}\right)\hat{F}_k(M,z)\right.}
    +\left.\mathrm{xf}_q \left(\frac{M}{\sqrt{s}}e^{-Y}\right)\mathrm{xf}_{\bar{q}}\left(\frac{M}{\sqrt{s}}e^{Y}\right)\hat{F}_k(M,-z)\right],
\end{equation}
\end{widetext}
which are defined correspondingly to
(\ref{eqn:PartonLevelXSecInNewVariables}). The plots  of the
$F_k(M,Y,z)$ $z$-dependence at fixed $M$, $Y$ are shown in Figures
A.1 -- A.7 of Appendix A. For small invariant masses ($M < 100-120$~GeV), the $F_3$ and $F_4$ functions are almost symmetric and 
therefore are suppressed in $A_{FB}$. This observation leads to idea that only the two first terms in (\ref{eqn:PartonXSecWithFactors}) are dominant
for the asymmetry. However,  such behaviour does not preserve for more  heavy $M$ bins ($M > 100-120$~GeV). As we see   in the figures A.5 -- A.7, 
in this case the $F_3$ and $F_4$ functions demonstrate behaviour which
significantly contribute to the asymmetry. So that the number of the unknown functions could not be reduced for heavy invariant masses. Nevertheless,
 the  $A_{F B}$  remains convenient observable because it is very sensitive to the the small changes of the coupling values everywhere. 
Thus, to analyze the
$A_{FB}$, we preserve in the cross-section $\frac{d^3\sigma}{dM\,dY\,dz}$   all the   terms entering \Ref{eqn:PartonXSecWithFactors}. 

 Next important  notice is that the CMS
detector has a finite acceptance and  only the leptons with $p_T >
p_0 = 20$~GeV can be detected. Therefore, to obtain the
cross-section of interest we have to integrate
the distributions  over $z$
in the  interval  $-z_0$ to $+z_0$, where
\begin{equation} \label{eqn:z0Definition}
    z_0 = \sqrt{1-4p^2_0/M^2}.
\end{equation}

\section{Estimation of  $Z'$ couplings}
The forward-backward asymmetry  is defined as
\begin{equation} \label{eqn:AFBDefinition}
    A_{FB} = \frac{\sigma_F - \sigma_B}{\sigma_F + \sigma_B},
\end{equation}
where
\begin{equation}
    \sigma_F = \int_{0}^{z_0} \frac{d\sigma}{dz}dz,~~ \sigma_B = \int_{-z_0}^{0} \frac{d\sigma}{dz}dz
\end{equation}
and $z_0$ is given in  (\ref{eqn:z0Definition}). Providing the notations
\begin{equation} \label{eqn:DeltaSigmaDefinitions}
    \Delta = \sigma_F - \sigma_B, ~~\Sigma = \sigma_F + \sigma_B,
\end{equation}
 we can rewrite (\ref{eqn:AFBDefinition}) in terms of the $Z'$ contributions,	
\begin{equation} \label{eqn:AFBWithZp}
\begin{split}
A_{FB}(M,Y) = \frac{\Delta(M,Y)}{\Sigma(M,Y)} = \\
 = \frac{\Delta^{SM} + \bar{a}^2 \Delta_1 + \bar{a}\bar{v}_l\Delta_2 + \bar{a}\bar{v}_u \Delta_{3} +  \bar{v}_l \bar{v}_u \Delta_{4}}{\Sigma^{SM} + \bar{a}^2 \Sigma_1 + \bar{a}\bar{v}_l\Sigma_2 + \bar{a}\bar{v}_u \Sigma_{3} + \bar{v}_l \bar{v}_u \Sigma_{4}},
\end{split}
\end{equation}
where, according to (\ref{eqn:HadronLevelZpFactors}),
\beao
    \Delta_k(M,Y) = \int_0^{z_0}{F_k(M,Y,z)\,dz}-\int_{-z_0}^0{F_k(M,Y,z)\,dz},\\\nn
    \Sigma_k(M,Y) = \int_0^{z_0}{F_k(M,Y,z)\,dz}+\int_{-z_0}^0{F_k(M,Y,z)\,dz}.\nn
\eeao
Expression (\ref{eqn:AFBWithZp}) is used for fitting the  $Z'$ parameters.

\begin{table}[h!]
    \caption{The CL intervals for the $Z'$ couplings} \label{tbl:ResultsCouplingsCLAreas}
    \centering
    \begin{tabular}{|c|c|c|}
        \hline
        \parbox[c]{0.8cm}{\vspace{2pt}$m_{Z'}$,\\GeV\vspace{2pt}} & 92\% CL boundaries, 7~TeV & 92\% CL boundaries, 8~TeV \\ \hline\hline
        \multirow{2}{*}{1200} & $\bar{a}^2=(1.5^{+36.5}_{-1.4})\times 10^{-5}$ & $\bar{a}^2=(1.3^{+20.8}_{-1.2})\times 10^{-5}$ \\ \cline{2-3}
        & $\bar{a}\bar{v}_l=(-0.4^{+3.8}_{-3.8})\times 10^{-5}$ & $\bar{a}\bar{v}_l=(-0.2^{+5.5}_{-14.2})\times 10^{-5}$ \\ \cline{2-3}
        & $\bar{a}\bar{v}_u=(3.4^{+4.2}_{-3.0})\times 10^{-3}$ & $\bar{a}\bar{v}_u=(3.4^{+1.7}_{-1.8})\times 10^{-3}$ \\ \hline\hline
        \multirow{2}{*}{3000} & $\bar{a}^2=(2.3^{+38.7}_{-1.2})\times 10^{-5}$ & $\bar{a}^2=(1.3^{+20.9}_{-1.2})\times 10^{-5}$ \\ \cline{2-3}
        & $\bar{a}\bar{v}_l=(-0.6^{+6.7}_{-0.8})\times 10^{-5}$ & $\bar{a}\bar{v}_l=(-0.2^{+5.5}_{-14.1})\times 10^{-5}$ \\ \cline{2-3}
        & $\bar{a}\bar{v}_u=(4.0^{+3.6}_{-3.6})\times 10^{-3}$ & $\bar{a}\bar{v}_u=(3.4^{+1.7}_{-1.8})\times 10^{-3}$ \\ \hline\hline
        \multirow{2}{*}{3500} & $\bar{a}^2=(2.4^{+38.6}_{-1.3})\times 10^{-5}$ & $\bar{a}^2=(1.3^{+20.9}_{-1.2})\times 10^{-5}$ \\ \cline{2-3}
        & $\bar{a}\bar{v}_l=(-0.6^{+6.8}_{-0.8})\times 10^{-5}$ & $\bar{a}\bar{v}_l=(-0.2^{+5.5}_{-14.1})\times 10^{-5}$ \\ \cline{2-3}
        & $\bar{a}\bar{v}_u=(4.0^{+3.6}_{-3.6})\times 10^{-3}$ & $\bar{a}\bar{v}_u=(3.4^{+1.7}_{-1.8})\times 10^{-3}$ \\ \hline\hline
        \multirow{2}{*}{4000} & $\bar{a}^2=(2.4^{+38.6}_{-1.3})\times 10^{-5}$ & $\bar{a}^2=(1.3^{+20.9}_{-1.2})\times 10^{-5}$ \\ \cline{2-3}
        & $\bar{a}\bar{v}_l=(-0.6^{+6.9}_{-0.8})\times 10^{-5}$ & $\bar{a}\bar{v}_l=(-0.2^{+5.5}_{-14.1})\times 10^{-5}$ \\ \cline{2-3}
        & $\bar{a}\bar{v}_u=(4.0^{+3.6}_{-3.6})\times 10^{-3}$ & $\bar{a}\bar{v}_u=(3.4^{+1.7}_{-1.8})\times 10^{-3}$ \\ \hline\hline
        \multirow{2}{*}{4500} & $\bar{a}^2=(2.4^{+38.6}_{-1.3})\times 10^{-5}$ & $\bar{a}^2=(1.3^{+20.9}_{-1.2})\times 10^{-5}$ \\ \cline{2-3}
        & $\bar{a}\bar{v}_l=(-0.6^{+6.8}_{-0.8})\times 10^{-5}$ & $\bar{a}\bar{v}_l=(-0.2^{+5.5}_{-14.1})\times 10^{-5}$ \\ \cline{2-3}
        & $\bar{a}\bar{v}_u=(4.0^{+3.6}_{-3.6})\times 10^{-3}$ & $\bar{a}\bar{v}_u=(3.4^{+1.7}_{-1.8})\times 10^{-3}$ \\ \hline
    \end{tabular}
\end{table}
\begin{figure}[h!]
    \caption{The 92\% CL area for the $Z'$ couplings: $(\bar{a}, \bar{v}_l)$ plane at $m_{Z'}=3$~TeV, $\bar{v}_u = 5\times 10^{-2}$}
    \centering
    \includegraphics{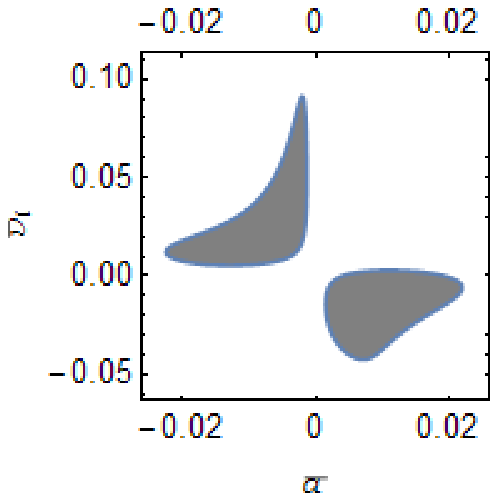}
    \label{fig:AVlCLArea}
\end{figure}
\begin{figure}[h!]
    \caption{The 92\% CL area for the $Z'$ couplings: $(\bar{a}, \bar{v}_q)$ plane at $m_{Z'}=3$~TeV, $\bar{v}_l = 5\times 10^{-4}$}
    \centering
    \includegraphics{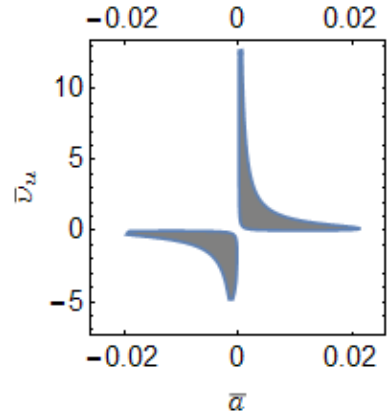}
    \label{fig:AVuCLArea}
\end{figure}
We calculate $\Sigma^{SM}$ by means of FEWZ~3 \cite{FEWZ} and  $A^{SM}_{FB}$, $\Delta^{SM}$, $\Delta_{1,2}$, and $\Sigma_{1,2}$ -- by using  Wolfram Mathematica 10 \cite{Mathematica}, FeynArts and FormCalc \cite{FeynArts}. Some of computations were fulfilled at the Dubna cluster HybriLIT \cite{Hybrilit}.  The accuracy of all these calculations is considered in detail in Discussion.

The results of the carried out calculations are presented in Table \ref{tbl:ResultsCouplingsCLAreas}. They demonstrate at almost $2 \sigma $ CL that the data on the $A_{FB}$ at $\sqrt{s} = 7$ and 8~TeV  are compatible with the $Z'$ existence. The estimates of all the $Z'$  couplings to the SM fermions  are obtained. The values of  parameters $\bar{a}\bar{v}_l$ and $\bar{a}\bar{v}_u$ are found as independent variables first in the literature. 
\section{Discussion}
We have analyzed the data on the $A_{FB}$ for the Drell-Yan
annihilation  process at the LHC presented by the CMS
collaborations for $\sqrt{s} = 7$ TeV \cite{CMS_AFB_7TeV} and 8~TeV \cite{CMS_AFB_8TeV} with the goal
of estimation in a model-independent approach the couplings of
the Abelian $Z'$ boson to the SM fermions. The investigation was
carried out within the effective Lagrangian \Ref{ZZplagr},
\Ref{ZZ'plagr}. As the important ingredient the 
 relations \Ref{grgav} were used. They essentially decreased
the number of couplings, which must be fitted, and modified
accordingly  the kinematics structure of the cross-sections. As a
result, the angular distribution of the theoretic cross-section
became uniquely  determined by this particle. It is important to
note that the relations  are satisfied at tree-level in all the
extended models investigated by the CMS and ATLAS \cite{ATLAS_HighMass_7TeV} -- \cite{CMS_ttbar_8TeV}
collaborations in the model-dependent approach. They also cover
other renormalizable models of Abelian $Z'$ \cite{Gulov2010}. Due to these
constraints, we  performed the three-parametric fit of the experimental data
and estimated the unknown $\bar{a}^2$, $\bar{a}\bar{v}_l$ and  $\bar{a}\bar{v}_u$
couplings for a number of $m_{Z'}$. 

The maximum likelihood method was applied. The QCD sector was evaluated with an NNLO accuracy, while the electroweak corrections were calculated up to NLO. This is a standard for the Drell-Yan production description at the LHC nowadays. The NLO effects are accounted for by means of an improved Born approximation (IBA). As it is known, the IBA absorbs the majority of NLO electroweak corrections. It is shown in \cite{Huber} its deviation from exact NLO calculations does not exceed 1-2\%. In the IBA approach, the coupling constants are replaced with the effective running couplings which are obtained from the one-loop expressions for the self-energy and vertex corrections. In fact, it means that we use an elastic scattering approximation but in all calculations we replace $\alpha_{em}(0)$ with $\alpha_{em}(m_Z)$. The PDF uncertainties were estimated by means of the standard formula (see, for example, \cite{MSTWPDF})
\begin{equation} \label{eqn:PDFUncertainty}
\Delta F = \frac{1}{2}\sqrt{\sum_{k=1}^n\left[F\left(S^+_k\right)-F\left(S^-_k\right)\right]^2},
\end{equation}
where $F$ is any quantity which depends on PDFs, $S^\pm_k$ are the PDF eigenvectors, and summation is performed over all the eigenvectors present in a given PDF set. In Table \ref{tbl:PDFEigenvectors}, we show the example for $\Sigma_k$ factors in the bin $86 \leq M \leq 96$, $0 \leq |Y| \leq 1$. From the presented results we see that the PDF uncertainty of the $Z'$ cross-sections does not exceed 3\%. Finally, considering all the discussed uncertainties as independent, we obtain that total theoretical uncertainty of the $A_{FB}$ predicted by (\ref{eqn:AFBWithZp}) is not larger than 5\%.

 The uncertainties
followed from  the statistical and the PDF errors were
calculated at $\sim2\sigma$ CL. It was concluded that
the $Z'$ existence is admitted by the data on $A_{FB}$
measured by the CMS at $\sqrt{s} = 7$ and 8~TeV. The $Z'$  signal (hint,
in fact) is non-zero at $92 \%$ CL. The obtained numerical
values for the $Z'$ coupling $\bar{v}^2_l$  are in an agreement with
the ones found  already for the LEP \cite{Gulov2010} and Tevatron
\cite{GulovKozhushkoModelIndependent} in a model-independent analysis where
other observables were proposed.

It is worth noting that the $\bar{a}\bar{v}_l$ coupling and $\bar{a}\bar{v}_u$ were estimated  directly for the first time.  In all other previous analysis only $\bar{v}^2_l$ could be estimated, while $\bar{a}\bar{v}_l$ was suppressed due to the process kinematics. Let us compare those values with our results. The calculation yields $\bar{v}^2_l<2.8\times 10^{-4}$ which is in agreement with $\bar{v}^2_l=(2.25^{+1.79}_{-2.07})\times 10^{-4}$ from \cite{Gulov2010} and $\bar{v}^2_l < 1.69 \times 10^{-4}$
from \cite{GulovKozhushkoModelIndependent}. Further, as we see from Table \ref{tbl:ResultsCouplingsCLAreas}, the experimental CMS data at 7 and 8 TeV, which were obtained with different precision, lead to the close values for estimated parameters.

It is essential that  the obtained  coupling values are weakly
dependent on the $Z'$ mass. It
is caused by the cross-section dependence on this parameter.
Really, the factors $\hat{F}_{qk}$ in
Eq.~\Ref{eqn:PartonXSecWithFactors} depend on the $m_{Z'}$ through
the $Z'$ propagator. This  is a denominator effect, which is small
at not close to the $Z'$ pole position energies.  On the contrary,
the couplings enter the cross-section through the numerator. Hence, the observables are much more sensitive to the coupling
variations.

Now let us turn back to Eq.~\Ref{MixingAngle}. The current limit on the $Z-Z'$ mixing angle from the global fit of the LEP data is about $|\theta_0| = 10^{-3}-10^{-4}$. We use this value to estimate the $m_{Z'}$. Due to \Ref{MixingAngle}, $\theta_0$ is expressed through $\bar{a}$ and $m_{Z'}$.
Since $\bar{a}$ is already derived, it is possible to obtain the $m_{Z'}$ limits which  satisfy the LEP restrictions on $\theta_0$. The optimistic estimation is $3 < m_{Z'} < 7-8$~TeV. The experimental  lower limit  was recently increased to the $ m_{Z'} > 3.5$ TeV  in the model-dependent analysis presented by the  CMS and ATLAS Collaborations.  The $Z'$ with such mass is possible to be detected in the future  LHC  experiments. Nevertheless in this case it will be difficult to distinguish the basic $Z'$ model. Therefore the model-independent description becomes an important instrument for investigating this problem. The obtained values of the couplings could be used in the $Z'$ model identifications  either at present or future colliders.

Finally we compare our results for $\bar{a}^2$  with those of in \cite{Gulov2010}, \cite{GulovKozhushkoModelIndependent},  where the data of the LEP and some LHC
experiments have been analyzed on the same principles as in the present paper. The
essential difference, however,  is  that in the former case it
was possible to introduce a one-parameter observable for
estimating the $\bar{a}^2$. The $\bar{a}\bar{v_l}$ contribution was excluded due to
more simple kinematics structure of the lepton cross-sections for the processes
$e^+ e^- \to \mu^+ \mu^- ( \tau^+ \tau^-)$.
 The  $\bar{a}^2$ found in \cite{Gulov2010} has the
value $\bar{a}^2 \leq 0.95\times 10^{-3}$ that differs from  our result $\bar{a}^2=(2.4^{+38.6}_{-1.3})\times 10^{-5}$. This is a universal parameter
related due to  \Ref{MixingAngle} with the $Z - Z'$ mixing angle, which was estimated at LEP experiments \cite{LEPThetaConstraints09}. On the contrary, the value of $\bar{a}^2$
found in \cite{GulovKozhushkoModelIndependent} is one order larger than obtained in Section 3. We could explain this discrepancy by the approximation for
the Drell-Yan process cross section used in \cite{GulovKozhushkoModelIndependent},
which is applicable at energies close to the resonance peak, only.
Possibly, this also depends on the data set and observables
introduced in the course of the analysis applied.

As conclusion we note that the applied model-independent approach can be used
for analyzing data of other experiments. In fact, at  the LHC  numerous data  on different processes have been accumulated.   Further improvements of the results 
 is expected from measurements fulfilled at run 2 of the LHC. 
So, it is of interest to investigate  these measurements by the applied method. Besides that, the $Z$ boson generation is  perspective
where the $Z-Z'$ mixing angle $\theta_0$ can be estimated and compared with the one obtained in the present paper. We left all these problems for the future.

\begin{acknowledgments}
 Authors are grateful to Alexey Gulov, Andrey Kozhushko and Alexander Pankov for useful discussions and suggestions.
\end{acknowledgments}

\appendix
\section{The Plots of the $Z'$ Factors}
\renewcommand\thefigure{A.\arabic{figure}}
\setcounter{figure}{0}
Below we present the behaviour of the  cross-section factors  introduced in (\ref{eqn:PartonXSecWithFactors}) and (\ref{eqn:HadronLevelZpFactors}). Here the functions $F_1$, $F_2$, $F_3$, and $F_4$ stand for the factors at $\bar{a}^2$, $\bar{a}\bar{v}_l$, $\bar{a}\bar{v}_u$, and $\bar{v}_l\bar{v}_u$, respectively.
\begin{figure}[h!]
    \caption{$F_{k}(M,Y,z)$ factors at $M=50$~GeV, $Y=1.25$}
    \centering
    \includegraphics{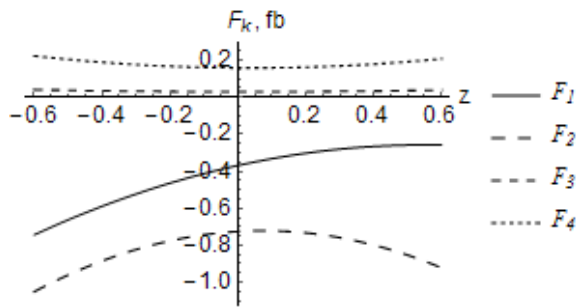}
    \label{fig:XSec1}
\end{figure}
\begin{figure}[h!]
    \caption{$F_{k}(M,Y,z)$ factors at $M=60$~GeV, $Y=1.25$}
    \centering
    \includegraphics{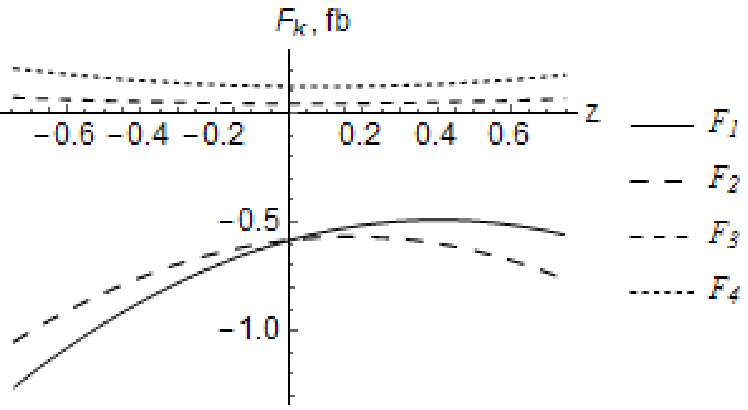}
    \label{fig:XSec2}
\end{figure}
\begin{figure}[h!]
    \caption{$F_{k}(M,Y,z)$ factors at $M=80$~GeV, $Y=1.25$}
    \centering
    \includegraphics{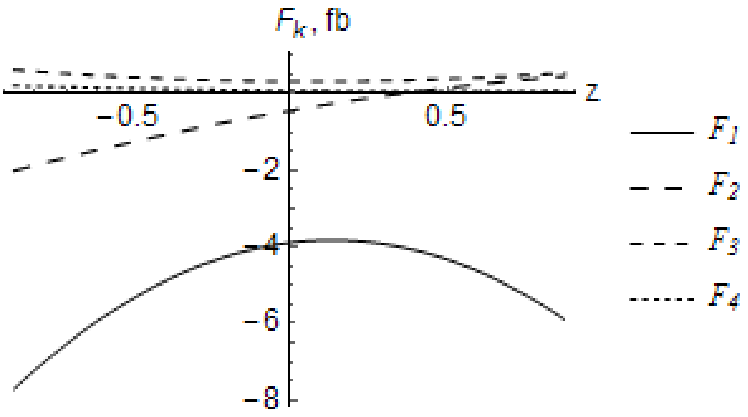}
    \label{fig:XSec3}
\end{figure}
\begin{figure}[h!]
    \caption{$F_{k}(M,Y,z)$ factors at $M=100$~GeV, $Y=1.25$}
    \centering
    \includegraphics{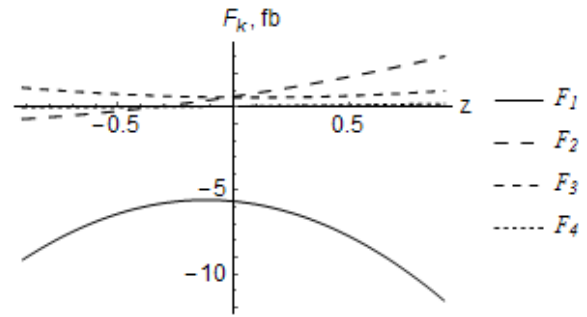}
    \label{fig:XSec4}
\end{figure}
\begin{figure}[h!]
    \caption{$F_{k}(M,Y,z)$ factors at $M=120$~GeV, $Y=1.25$}
    \centering
    \includegraphics{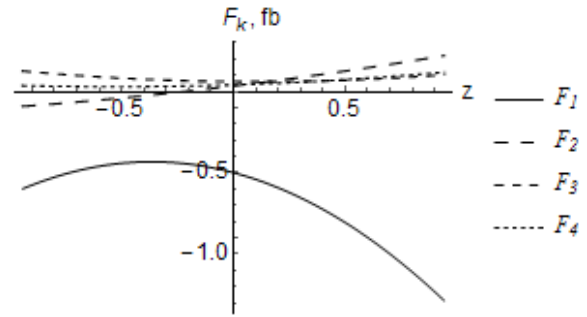}
    \label{fig:XSec5}
\end{figure}
\begin{figure}[h!]
    \caption{$F_{k}(M,Y,z)$ factors at $M=200$~GeV, $Y=1.25$}
    \centering
    \includegraphics{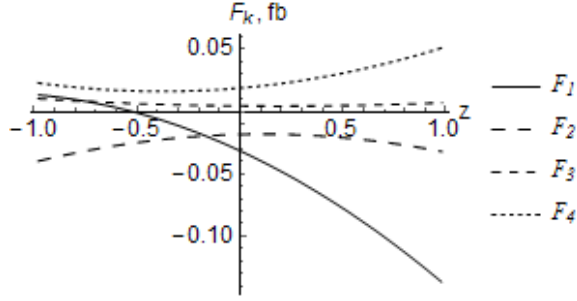}
    \label{fig:XSec6}
\end{figure}
\begin{figure}[h!]
    \caption{$F_{k}(M,Y,z)$ factors at $M=400$~GeV, $Y=1.25$}
    \centering
    \includegraphics{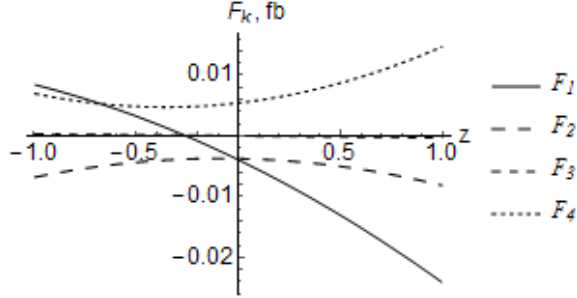}
    \label{fig:XSec7}
\end{figure}

\section{On the Relation between the $Z'$ Couplings}
Below we adduce an information on the
 derivation of  mixing angle
\Ref{MixingAngle} and correlations \Ref{grgav}. As it was noted in Sect. 2, these correlations
are proper to renormalizable  models containing the Abelian $Z'$
boson. They have been obtained in \cite{Gulov00a},
\cite{Gulov01} by using two different procedures (see for  details
\cite{Gulov2010}). General idea of the first approach is mentioned
in the main text. The second way is based on principle of gauge
invariance with respect to the $\tilde U(1)$ transformations
\cite{Gulov01}.

The most general effective Lagrangian describing the $Z'$
interactions
 with the SM fields and preserving the $SU(2)_L \times
U(1)_Y \times \tilde U(1)$ gauge group reads \cite{Cvetic},
\cite{Sirlin},
\bea \label{Leffective}
\begin{split}
\mathcal{L} = \frac{1}{2} \left|\left( i
D^{ew,\phi}_\mu + \tilde g \tilde Y_\phi Z'_{0\mu}\right) \phi \right|^2 \\
+ \sum \limits_{f_L} \bar f_L \left( i D^{ew,L}_\mu + \tilde
g \tilde Y_{f}^L Z'_{0\mu}\right)\gamma^\mu f_L \\
+ \sum
\limits_{f_R} \bar f_R \left( i D^{ew,R}_\mu + \tilde g \tilde
Y^R_f Z'_{0\mu}\right) \gamma^\mu f_R,
\end{split}
 \eea
where the summation over all the SM left-handed doublets, $f_L$, and the SM right-handed singlets, $f_R,$ is assumed and  $D_\mu^{ew,L} = \partial_\mu - \frac{i g}{2} \sigma^a
A^a_\mu - \frac{i g'}{2} Y_{f,L} B_\mu$ is the standard model
 covariant derivative with the values of the hypercharges: $Y_\phi = 1$ , $Y_{f,L} = \frac{1}{3}$
and $D^{ew,R}_\mu = \partial_\mu - i g' Q_f B_\mu$, $Q_f$ is the
fermion charge in the positron charge units, $\sigma^a$ are the
Pauli matrices. The values of the dimensionless constants $\tilde
Y_\phi, \tilde Y^L_{f}, \tilde Y^R_{f}$ depend on a particular
model and here are considered as arbitrary numbers.

The masses of the SM particles are generated by the spontaneous
breaking of the $SU(2)_L \times U(1)_Y \to U(1)_{em}$ symmetry due
to the non-zero vacuum value of the scalar doublet.  Hence, the mass
eigenstates of the vector bosons are appeared to be shifted from the original fields
$A^a_\mu, B_\mu, Z'_{0\mu}$ because the corresponding mass matrix became nondiagonal.
  Physical fields $A_\mu, Z_\mu, Z'_{
\mu}$ are obtained by the orthogonal transformation:
\bea \label{fields} B_\mu &=& A_\mu c_W - (Z_\mu c_0 - Z'_\mu s_0)
s_W \\ \nn A^3_\mu &=& A_\mu s_W + (Z_\mu c_0 - Z'_\mu s_0) c_W \\
\nn Z'_{0\mu} &=& Z_\mu s_0 + Z'_\mu c_0, \eea
where $c_W = \cos \theta_W, s_W = \sin \theta_W$ and the SM value
of the Weinberg angle $\tan \theta_W = g'/g$. Whereas $c_0 =
\cos \theta_0, s_0 = \sin \theta_0$ denote the cosine and sine of
the mixing angle $\theta_0$ relating the physical states $Z_\mu,
Z'_\mu$ to the massive neutral components of the $SU(2)_L \times
U(1)_Y \times \tilde U(1)$  gauge fields. The value of the
$\theta_0$ can be determined from the relation
\be \label{angle0} \tan^2 \theta_0 = \frac{m^2_W/c^2_W -
m^2_Z}{m^2_{Z'} - m^2_W/c^2_W}, \ee
(see also \cite{Sirlin}) expressing it through the masses of
physical states, which appeared after the orthogonalization, and the SM Weinberg agle. The
difference in the numerator of the r.h.s. is positive and
completely determined by the $Z'$ coupling to the scalar field
doublet \cite{Gulov01}. After the diagonalization, the masses of physical states are given by 
\bea \label{masses} m^2_A &=& 0, \\ \nn
          m^2_Z&=& m_W^2 c_W^{-2} \left(1 - \frac{4 \tilde g^2 \tilde Y^2_{\phi}}{g^2} \frac{m^2_W}{  m^2_{Z'} -  m_W^2 c_W^{-2}}\right), \\ \nn
    m^2_{Z'}&=& m^2_{Z'_0} + ( m_W^2 c_W^{-2} -  m^2_Z ) +   \frac{4 \tilde g^2 \tilde Y^2_{\phi}}{g^2} m^2_W.
\eea
Here, $ m_{Z'_0}$ is the mass of the $Z'$ before diagonalization. This value is not specified and is a free parameter. As we see, the mass $m_Z$ differs
from the SM value $m_W/c_W$ by a small quantity of the order $\sim m^2_W/m^2_{Z'}$. So the mixing angle \Ref{angle0}  is also small, $\theta_0 \sim m^2_W/m^2_{Z'}$
%
%
Using \Ref{fields} the Lagrangian of the model can be expressed in
terms of the physical fields. The $\theta_0$-dependent terms
generate new interactions originally absent in \Ref{Leffective}. Here it worth reminding that in calculations carried out we used the SM value of the Weinberg angle $\tan \theta_W = g'/g$. 

To derive the correlations \Ref{grgav} we require the Yukawa terms
of the SM to be invariant with respect to the $\tilde U(1)$ gauge
symmetry. This condition is fulfilled if the relation holds
\be \label{relation} \tilde Y^R_f = \tilde Y^L_f + 2 T^3_f ~\tilde
Y_\phi. \ee
Introducing the $Z'$ interaction constants with the vector and
axial-vector  currents of fermions $v_{Z'}^f = \frac{\tilde{g}}{2}(\tilde
Y^L_f  + \tilde Y^R_f),~ a_{Z'}^f = \frac{\tilde{g}}{2}(\tilde Y^L_f  -
\tilde Y^R_f)   $ we can rewrite \Ref{relation} in the form
\Ref{grgav}. These correlations also hold in Two-Higgs-Doublets SM
\cite{Gulov00a}, \cite{Gulov2010}.

 The relation \Ref{relation}  is just  as in the SM for the given  proper values of the
hypercharges $Y^R_f, Y^L_f, Y_\phi.$ In the extended models, the
originally independent parameters  $\tilde Y^R_f, \tilde Y^L_f,
\tilde Y_\phi$ have to be connected ones. The fermion and the
scalar sectors of the $Z'$  physics are  correlated. In particular, the mixing angle is simply related with the universal  axial vector coupling $a_f^2$.

\section{ PDF uncertainties}

In this Appendix, we present table which illustrates calculation of the PDF uncertainties with Eq. (\ref{eqn:PDFUncertainty}). It shows the $Z'$ factors $\Sigma_k$ integrated over the bin $86\leq M \leq 96$, $0 \leq |Y| \leq 1$ with some of the PDF eigenvectors:
$$\Sigma_k = \int_{86~GeV}^{96~GeV}{dM\int_{-1}^{1}dY\,\Sigma_k(M,Y)}.$$
$\Sigma_k(M,Y)$ are defined in (\ref{eqn:DeltaSigmaDefinitions}), and $\langle\Sigma\rangle$ means the central values.

\begin{table}[h!]
    \caption{The $Z'$ factors calculated with different PDF eigenvectors and integrated over $86\leq M \leq 96$, $0 \leq |Y| \leq 1$} \label{tbl:PDFEigenvectors}
    \centering

	\begin{tabular}{|c|c|c|c|c|}
	 \hline
	 & $\Sigma_1$, pb & $\Sigma_2$, pb & $\Sigma_3$, pb & $\Sigma_4$, pb \\
	 \hline\hline
 1 & $-3366.02$ & $200.44$ & $205.64$ & $1.96$ \\
 2 & $-3364.16$ & $200.31$ & $205.59$ & $1.96$ \\
 3 & $-3364.55$ & $200.50$ & $207.17$ & $1.95$ \\
 4 & $-3365.51$ & $200.27$ & $204.30$ & $1.97$ \\
 5 & $-3373.87$ & $200.62$ & $207.58$ & $1.98$ \\
 6 & $-3358.97$ & $200.21$ & $204.23$ & $1.96$ \\
 7 & $-3369.25$ & $200.91$ & $206.24$ & $1.97$ \\
 8 & $-3362.64$ & $200.06$ & $205.23$ & $1.96$ \\
 9 & $-3414.86$ & $203.46$ & $208.02$ & $2.00$ \\
 10 & $-3323.79$ & $197.83$ & $203.62$ & $1.94$ \\
 11 & $-3362.33$ & $200.58$ & $210.16$ & $1.94$ \\
 12 & $-3365.90$ & $200.28$ & $203.90$ & $1.97$ \\
 13 & $-3350.05$ & $199.40$ & $204.09$ & $1.95$ \\
 14 & $-3374.68$ & $201.01$ & $206.62$ & $1.97$ \\
 15 & $-3359.04$ & $200.06$ & $205.05$ & $1.96$ \\
 16 & $-3354.31$ & $199.71$ & $205.13$ & $1.96$ \\
 17 & $-3366.81$ & $200.26$ & $206.29$ & $1.97$ \\
 18 & $-3361.63$ & $200.50$ & $204.53$ & $1.96$ \\
 19 & $-3360.79$ & $200.18$ & $205.07$ & $1.96$ \\
 20 & $-3345.18$ & $199.14$ & $204.75$ & $1.95$ \\ \hline\hline
 $\langle\Sigma\rangle$, pb & $-3365.09$ & $200.37$ & $205.61$ & $1.96$ \\ \hline
 $\frac{\Delta\Sigma}{\langle\Sigma\rangle}$, \% & $1.6$ & $1.4$ & $2.6$ & $2.1$ \\ \hline
	\end{tabular}
\end{table}

\bibliography{A_FB}

\end{document}